\documentclass[twocolumn,showpacs,preprintnumbers,amsmath,amssymb]{revtex4}

\usepackage{graphicx}
\usepackage{dcolumn}
\usepackage{bm}

\begin{document}


\title{Squeezing and entanglement in continuous variable systems\\}

\author{Yun-Jie Xia$^{1,2}$}
\email{yjxia@qfnu.edu.cn}
\author{Guang-Can Guo$^{1}$}
\email{gcguo@ustc.edu.cn} \affiliation {
\centerline{ $^{1}$Key Laboratory of Quantum Information,
University of
Science and Technology of China,Chinese Academy of Sciences,\\
            }
\centerline{
Hefei, Anhui, China, 230026.\\
           }
\centerline{
$^{2}$Department of Physics, Qufu Normal University, Qufu, Shandong, China, 273165.\\
           }
}

\begin{abstract}
Based on total variance of a pair of Einstein-Podolsky-Rosen (EPR)
type operators, the generalized EPR entangled states in continuous
variable systems are defined. We show that such entangled states
must correspond with two-mode squeezing states whether these
states are Gaussian or not and whether they are pure or not. With
help of the relation between the total variance and the
entanglement, the degree of such entanglement is also defined.
Through analyzing some specific cases, we see that this method is
very convenient and easy in practical application. In addition, an
entangled state with no squeezing is studied, which reveals that
there certainly exist something unknown about entanglement in
continuous variable systems. \\
\end{abstract}

\maketitle
\section{Introduction}
\renewcommand{\theequation}{\arabic{section}.\arabic{equation}}

Quantum entanglement is the most significant and oddest trait in
quantum mechanics and quantum information{\cite{entan1,entan2}}.
Entanglement plays a crucial role in quantum information
processing, such as quantum teleportation {\cite{teleport}},
entanglement swapping {\cite{Ekert1}}, dense coding
{\cite{coding1}}, quantum cryptography {\cite{Ekert2}} and quantum
computation {\cite{Ekert3}}. After the first experiments on
quantum teleportation {\cite{Kimble1}} and other quantum
information processes using two-mode squeezing states
{\cite{Vaidman,Kimble2}}, continuous variable systems have aroused
great interest in the separability properties. So far, most of
theoretical and experimental work has focused on the entanglement
properties of Gaussian states. For Gaussian states, the necessary
and sufficient inseparability criterion had been fully
developed{\cite{Duan1,Simon}. Recently, continuous-variable
quantum teleportation through lossy channels{\cite{Welsch}},
quantum homogenization with linear optical elements{\cite{Buzek}}
and continuous-variable Werner
state {\cite{Mista}} for Gaussian sates have been studied.\\
\indent Generally speaking, it is difficult to determine whether a
state is an entangled sate. Although this problem has been solved
in mathematics {\cite{Horo1}}, it is not easy to make judgement
for most of practical physics systems. In this paper, we consider
the more generalized case that a state may not be necessarily a
Gaussian state. Using the total variance of a pair of
Einstein-Podolsky-Rosen type operators that is introduced by Duan
et al. {\cite{Duan1}}, we obtain the sufficient condition for
entanglement in continuous variable systems. This kind of
entangled states are regarded as generalized EPR entangled states
(GEES). We also prove that a state must be two-mode squeezed state
if the state is a GEES whether it is Gaussian or not, and whether
it is pure or not. From this result, the degree of entanglement
{\cite{Wootter1}} can be defined for such kind of entangled states
and some examples are discussed in detail. It is clearly seen that
the sufficient condition is very convenient and easy to judge
whether a state is an entangled state.\\
\indent This paper is organized as follows. In Section
{\ref{GEES}}, we derive the sufficient condition for entanglement.
In section {\ref{DEG}}, we define the degree of entanglement. Some
examples are discussed in detail in section { \ref{SAMPLE}}, in
which we prove that there exists an entangled state with no
two-mode squeezing.
\\
 \section {The total variance of EPR-like operators and
 entanglement}\label{GEES}
\setcounter{equation}{0}
For two-particle continuous variable systems, the maximum
entangled state may be expressed as a common eigenstate of EPR
type operators{\cite{Einstein}}
\begin{eqnarray}\label{epr1}
\hat{X}_{+}=\hat{x}_{1}+\hat{x}_{2}\text{\quad and \quad}
\hat{P}_{-}=\hat{p}_{1}-\hat{p}_{2}
\end{eqnarray}
or another pair of operators
\begin{eqnarray}\label{epr2}
\hat{X}_{-}=\hat{x}_{1}-\hat{x}_{2}\text{\quad and \quad}
\hat{P}_{+}=\hat{p}_{1}+\hat{p}_{2}
\end{eqnarray}
 where $x_{i}$ and $p_{i}$ (i=1,2) are the position and momentum operators for $i$-th particle.
Following Ref.{\cite{Duan1}}, we introduce the EPR-like operators
\begin{eqnarray}\label{eprlike}
\hat{u}=|c|\hat{x}_{1}+\frac{1}{c}\hat{x}_{2}\\
\hat{v}=|c|\hat{p}_{1}-\frac{1}{c}\hat{p}_{2}
\end{eqnarray}
If $c=\pm 1$, above operators are reduced to the standard EPR type
operators as expressed in (\ref{epr1}) and (\ref{epr2}).\\
\indent In {\cite{Duan1}}, it has been proved if a state is
separable, then the total variance of EPR-like operators $\hat{u}$
and $\hat{v}$ will satisfy the inequality
\begin{eqnarray}\label{sep1}
\Delta\hat{u}^{2}+\Delta\hat{v}^{2}\geq c^{2}+\frac{1}{c^{2}}
\end{eqnarray}
\indent We let the position and momentum operators be rewritten in
terms of a pair of boson operators $\hat{a}_{i}$(annihilation
operators) and $\hat{a}^{\dagger}_{i}$(creation operators)
\begin{eqnarray}\label{boson}
\hat{a}_{i}=\frac{\hat{x}_{i}+i\hat{p}_{i}}{\sqrt{2}} \text{\qquad
and \qquad}
\hat{a}^{\dagger}_{i}=\frac{\hat{x}_{i}-i\hat{p}_{i}}{\sqrt{2}}
\end{eqnarray}
After some algebra, the total variance of EPR-like operators is
expressed as
\begin{eqnarray}\label{sep2}
\Delta\hat{u}^{2}+\Delta\hat{v}^{2}=c^{2}+\frac{1}{c^{2}}+2\langle\hat{T}\rangle
\end{eqnarray}
where
\begin{eqnarray}\label{test1}
\hat{T}=\left(\hat{Z}^{\dagger}-\langle\hat{Z}^{\dagger}\rangle\right)
\left(\hat{Z}-\langle\hat{Z}\rangle\right)-\frac{1}{c^{2}}
\end{eqnarray}
and
\begin{eqnarray}\label{test2}
\hat{Z}=|c|\hat{a}^{\dagger}_{1}+\frac{1}{c}\hat{a}_{2}
\end{eqnarray}
\indent It is well-known that displacement operation does not
change the entanglement of a state. Without loss of generality, we
can take $\bar{x}_{i}=\bar{p}_{i}=0(i=1,2)$, then
\begin{eqnarray}\label{test3}
\hat{T}=c^{2}\hat{a}^{\dagger}_{1}\hat{a}_{1}+\frac{1}{c^{2}}\hat{a}^{\dagger}_{2}
\hat{a}_{2} +\frac{c}{|c|}\left(\hat{a}_{1}\hat{a}_{2}
+\hat{a}^{\dagger}_{1}\hat{a}^{\dagger}_{2} \right)
\end{eqnarray}
Clearly, the expected value of $\hat{T}$ in any separable state
$\sigma$(density matrix) satisfies
\begin{eqnarray}\label{test4}
Tr(\hat{T}\sigma)\geq 0
\end{eqnarray}
According to inseparable theorem in {\cite{Horo1}}, if
\begin{eqnarray}\label{test5}
Tr(\hat{T}\rho)< 0
\end{eqnarray}
for a given state $\rho$, then state $\rho$ must be inseparable.
So we can call $\hat{T}$ as a test operator. If operator $\hat{T}$
satisfies Eq. (\ref{test5}), the state of the two-particle
continuous variable system is called  "generalized EPR-entangled
state"(GEES). If GEES exists in this system, it is easy to see
that there must be two-mode squeezing. Combining Eq.(\ref{sep2})
with (\ref{test5}) then we obtain
\begin{eqnarray}\label{squeez1}
\Delta\hat{u}^{2}<\frac{1}{2}\left(c^{2}+\frac{1}{c^{2}}\right)
\text{\quad or \quad}
\Delta\hat{v}^{2}<\frac{1}{2}\left(c^{2}+\frac{1}{c^{2}}\right)
\end{eqnarray}
This is just the squeezing condition of EPR-like operators
$\hat{u}$ and $\hat{v}$. So we conclude that there must be
two-mode squeezing if GEES exists in this system.\\
\indent It is noticeable that the parameter $c$ is not arbitrary
but dependent on the given state{\cite{Duan1}}. When we
investigate the entanglement of a given state, we always hope that
the variance of EPR-like operator is as small as possible. From
(\ref{test3}), the expected value of test operator $\hat{T}$ will
reach the smallest under the condition
\begin{eqnarray}\label{constantc}
c^{2}=\sqrt{\frac{n_{2}}{n_{1}}}
\end{eqnarray}
when the mean number $n_{1}$($n_{2}$) of photon for each mode is
fixed. Obviously, constant $c$ in EPR-like operators is determined
by the mean number of photon. When the mean photon numbers of two
modes is equal, it is allowed to let $c=\pm 1$. In practical
application, we can first find $c$ by the mean photon numbers
$n_{1}$($n_{2}$) conveniently.\\
\indent As to the sign of $c$, it depend on the expected value of
operator $\hat{a}_{1}\hat{a}_{2}
+\hat{a}^{\dagger}_{1}\hat{a}^{\dagger}_{2}$
\begin{eqnarray}\label{tleast1}
\delta=\langle\hat{a}_{1}\hat{a}_{2}
+\hat{a}^{\dagger}_{1}\hat{a}^{\dagger}_{2}\rangle
\end{eqnarray}
The sign of $c$ is in opposition to that of $\delta$.\\
\indent From the above analysis we finally get the condition for
GEES
\begin{eqnarray}\label{xia}
|\delta|>2\sqrt{n_{1}n_{2}}
\end{eqnarray}
It is obvious that this criterion is very simple and feasible in
practical application.\\
\indent Because the constant $c$ may be positive or negative, the
entanglement takes on two different types. One is near the common
eigenstate of total position operator and relative momentum
operator, the other is near the common eigenstate of total
momentum operator and relative position operator.\\
\indent In this section, we do not require that the related state
is pure or mixed, and it is Gaussian or not. The entanglement
condition (\ref{xia}) is held for any GEES.\\
\indent Nevertheless, the squeezing of EPR-like operators is only
the necessary condition for GEES. There must be some other type of
entangled states in continuous variable systems. We will introduce
an entangled state in which there is no squeezing in Sec.
{\ref{SAMPLE}}.
\section{Degree of Entanglement}\label{DEG}
\setcounter{equation}{0} \indent To find a well justified and
mathematically tractable measure of entanglement is likely to be
of value in number of areas of research, including the study of
decoherence in quantum computers{\cite{Divincenzo}}, and the
evaluation of quantum cryptographic schemes{\cite{Gisin}}. About
the degree of entanglement, the entanglement of formation is a
creative idea. Wootters{\cite{Wootter1}} first studied the
analytical expression of entanglement of formation for two-qubit
states. Later the entanglement of formation was also determined
for highly symmetric states{\cite{Terhal,Werner}}. In the above
section, the total variance of EPR-like operator is certainly
related to the entanglement. From (\ref{sep2}), the total variance
of EPR type operators is zero for the maximum entangled state.
Given a state $\rho$, if we define the "concurrence"(Here we
borrow the concept introduced by Wootter{\cite{Wootter1}})as
\begin{eqnarray}\label{conc1}
\tau_{\rho}=\text{min}\left[\frac{\Delta\hat{u}^{2}+\Delta\hat{v}^{2}}{c^{2}+1/c^{2}},1\right]
\end{eqnarray}
its value region is in $(0,1]$. Now we can define the degree of
entanglement as the following form
\begin{eqnarray}\label{conc2}
E(\tau_{\rho})=h\left(\frac{1+\sqrt{1-\tau^{2}_{\rho}}}{2}\right)\nonumber\\
h(x)=-x\log_{2}x-(1-x)\log_{2}(1-x)
\end{eqnarray}
The entanglement decreases along with increase of $\tau_{\rho}$,
so this function is a monotony convex in $\tau 's$ value
region{\cite{Wootter1}} and can be regarded as a measure of GEES.
For example, the two-mode squeezed state{\cite{Milburn}}
\begin{eqnarray}\label{twoentang}
|\psi\rangle=e^{r(\hat{a}_{1}^{\dagger}\hat{a}_{2}^{\dagger}-\hat{a}_{1}\hat{a}_{1})}|00\rangle
\end{eqnarray}
its "concurrence" is
\begin{eqnarray}\label{tconc3}
\tau_{\psi}=e^{-2r}<1
\end{eqnarray}
which means that the two-mode squeezed pure state is always
entangled.
\section {Some examples and discussions}\label{SAMPLE}
\setcounter{equation}{0} \indent In this section, we will analyze
in detail some examples about two-particle continuous variable
systems. Of these examples, one is the standard EPR-type entangled
state(such as Gaussian state with equal mean photon number and
minimum-correlation mixed state), and the other is an entangled
state but not EPR-like entanglement, in which there does not exist
any squeezing. From these examples, it is seen that there are
likely a lot more about entanglement to be studied in continuous
variable systems.\\
\subsection{Gaussian state with equal mean photon numbers}
The Gaussian state with equal mean photon numbers for two modes is
expressed by $Q$ function
\begin{eqnarray}\label{santos1}
Q(\alpha,\beta)&=&(1-x^{2})K^{2}\text{exp}\left\{
-K\left[|\alpha|^{2}+|\beta|^{2}\right.\right.\nonumber\\
&&\left.\left.+x(\alpha\beta+\alpha^{*}\beta^{*})\right]\right\}\nonumber\\
K&=&\frac{1}{(n+1)(1-x^{2})},n\ge 0, |x|<1
\end{eqnarray}
where $n$ is the mean number of photon per mode, and $x$ is a
correlation parameter. The inseparability of this state had been
studied in Ref.{\cite{Santos}} using Simon's
criterion{\cite{Simon}}. Now we will study it by (\ref{sep2}).
With help of integration
\begin{eqnarray}\label{integ}
\int \frac{d^{2}z}{\pi}e^{-\gamma|z|^{2}+sz+tz^{*}}
=\frac{1}{\gamma}e^{\frac{st}{\gamma}},\text{\qquad $\gamma>0$}
\end{eqnarray}
it is easy to obtain
\begin{eqnarray}\label{santos2}
\langle\hat{a}_{1}\hat{a}_{2}
+\hat{a}^{\dagger}_{1}\hat{a}^{\dagger}_{2}\rangle =-2x(n+1)
\end{eqnarray}
Because the mean photon numbers for two mode are equal, the
parameter $c\!=+1 \ \text{or} -1$ is dependent on the sign of
correlation constant $x$. The expected value of test operator
$\hat{T}$ is
\begin{eqnarray}\label{santos3}
\langle\hat{T}\rangle=2n\mp2x(n+1)
\end{eqnarray}
The entangled state corresponds to $\langle\hat{T}\rangle<0$,
which leads to
\begin{eqnarray}\label{santos4}
\frac{n}{n+1}<|x|\le\sqrt{\frac{n}{n+1}}
\end{eqnarray}
This is agreement with Ref.{\cite{Santos}} and show that this
method is simpler and easier than Simon's criterion. The limit for
$x$ in the right side of above equation comes from the requirement
that the $Q$ function in (\ref{santos1}) is not valid for all
values of $x$ {\cite{Santos}}.
\subsection{The minimum-correlation mixed state}
The minimum-correlation state{\cite{Wolf}} for the pair of
operators $\hat{X}_{+}$ and $\hat{P}_{+}$ is expressed in the
form{\cite{Agarwal}}
\begin{eqnarray}\label{agarwal}
Q(\alpha,\beta)&=&(1-2d)\text{exp}\left\{
-(1-d)\left[\gamma|\alpha|^{2}+\frac{1}{\gamma}|\beta|^{2}\right]\right\}\nonumber\\
&&\times\text{exp}\left[-d(\alpha\beta+\alpha^{*}\beta^{*})\right].
\end{eqnarray}
where $\gamma=\tanh(r)$ is the squeezing parameter. The
requirement that $Q$ function converge for large values of
$|\alpha|$ and $|\beta|$ gives the condition $d\le 1/2$.\\
\indent The expected values of related operators are given by the
expressions
\begin{subequations}
\label{agarwal2}
\begin{equation}
n_{1}=\frac{1-d}{\gamma(1-2d)}-1
\end{equation}
\begin{eqnarray}
n_{2}=\frac{\gamma(1-d)}{1-2d}-1
\end{eqnarray}
\begin{eqnarray}
\langle\hat{a}_{1}\hat{a}_{2}
+\hat{a}^{\dagger}_{1}\hat{a}^{\dagger}_{2}\rangle
=-\frac{2d}{1-2d}
\end{eqnarray}
\end{subequations}
On substituting from Eq.(\ref{agarwal2}) into Eq.(\ref{constantc})
and Eq.(\ref{test3}), we obtain the expression
\begin{eqnarray}\label{agarwal3}
\langle\hat{T}\rangle=-\left[\tanh(r)+\coth(r)-2\right]\frac{1-d}{1-2d}
\end{eqnarray}
This expected value is negative for all $r>0$, which means
entanglement always exists in the state.\\
\indent It is worth pointing out that the total variance of
standard EPR-type operators does not satisfy the entanglement
condition Eq.(\ref{test5}). In fact, one can derive the following
result
\begin{eqnarray}\label{argwarl4}
\langle\hat{T}\rangle|_{c=1}=\left[\tanh(r)+\coth(r)-2
\right]\frac{1-d}{1-2d}\ge 0
\end{eqnarray}
for $r>0$. This result shows that there does not  exist squeezing
of standard EPR-type operators in the minimum correlation state.
But the state certainly exhibits squeezing of EPR-like operators
with $c^{2}=\sqrt{n_{2}/n_{1}}$. So GEES corresponds with the
squeezing of standard EPR-type operators only in some special case
and the corresponding is held only when the mean photon number per
mode are equal. It is very important to choose the constant $c$
properly when we judge whether a state is an entangled state.
\subsection{The entangled coherent states}
The entangled coherent states{\cite{Sanders}} are other important
continuous variable entangled states. Although the coherence is
degraded when these states are embedded in the environment, its
entanglement is not totaly lost. Much possible application to
quantum information processing has been studied utilizing
entangled states{\cite{ECS}}. The entangled coherent states can be
defined as
\begin{eqnarray}\label{evenodd}
|\psi^{\pm}\rangle=N_{\pm}(|\alpha_{1}\rangle|\alpha_{2}\rangle\pm|-\alpha_{1}\rangle|-\alpha_{2}\rangle)
\end{eqnarray}
where $|\alpha_{i}\rangle(i=1,2)$ is coherent state for $i$-th
mode, $N_{\pm}=1/\sqrt{2[1\pm
e^{-2(|\alpha_{1}|^{2}+|\alpha_{2}|^{2})}]}$ are the normalization
constants. In the following discussion, without loss of
generality, we let $\alpha_{1}=\alpha_{2}$. The above two states
are certainly entangled states for two-mode fields. From the
following results
\begin{subequations}\label{evenodd1}
\begin{eqnarray}
\langle \psi^{+}|\psi^{-}\rangle=0
\end{eqnarray}
\begin{eqnarray}
\hat{a}_{1}\hat{a}_{2}|\psi^{\pm}\rangle=\hat{a}_{1}^{2}|\psi^{\pm}\rangle
=\hat{a}_{2}^{2}|\psi^{\pm}\rangle=\alpha^{2}|\psi^{\pm}\rangle
\end{eqnarray}
\begin{eqnarray}
\langle\psi^{\pm}|\hat{a}_{1}|\psi^{\pm}\rangle =
\langle\psi^{\pm}|\hat{a}_{2}|\psi^{\pm}\rangle=0
\end{eqnarray}
\begin{eqnarray}
n_{1}^{\pm}=n_{2}^{\pm}=|\alpha|^{2}\frac{N_{\pm}^{2}}{N_{\mp}^{2}}
\end{eqnarray}
\end{subequations}
one can obtain the expected values of the test operator
\begin{eqnarray}\label{evenodd2}
\langle\hat{T}\rangle_{+}&=&2R^{2}[\tanh(2R^{2})-|\cos2\theta|]
\end{eqnarray}
\begin{eqnarray}\label{evenodd3}
\langle\hat{T}\rangle_{-}&=&2R^{2}[\coth(2R^{2})-|\cos2\theta|]
\end{eqnarray}
where $\alpha=Re^{i\theta}$. Because $\langle\hat{T}\rangle_{-}$
is always positive for any value of parameter $\alpha$,
$|\psi^{-}\rangle$ is not a GEES. $|\psi^{+}\rangle$ is still a
GEES if $\alpha$ is chosen properly. \\
\indent It is easy to show that $|\psi^{+}\rangle$ can exhibit
squeezing of two-mode field under suitable condition. But there
does not exist any squeezing of two-mode field amplitude in the
state $|\psi^{-}\rangle$. Although GEES certainly leads to the
squeezing of EPR-like operator, the squeezing exits only in GEES
within our present knowledge. Other type of entangled states in
this system, such as $|\psi^{-}\rangle$ and $|\psi^{+}\rangle$
under some conditions, exhibit no squeezing. These kinds of
entanglement
have not been studied about their essence so far.\\
\indent From the above discussion we can conclude that GEES is the
only one kind of entanglement for two-particle continuous variable
systems. Other types of the entanglement in this system, such as
$|\psi^{\pm}\rangle$, are necessarily to be studied deeply.
Whether this kind of entanglement corresponds with some
nonclassical effects of two-mode fields and the entanglement
degree of such entangled states are all worthy of studying later.
It is reasonable to believe that there is much more to be studied
for this system although we noticed that some progress about the
entanglement measure for continuous variable
system has been made{\cite{Parker,{Lu}}.\\
\begin{acknowledgments}
We thank Lu-Ming Duan and Zheng-Wei Zhou for helpful discussions.
This work was supported by the National Fundamental Research
Program (2001CB309300), National Natural Science Foundation of
China, and the Innovation funds from the Chinese Academy of
Sciences.
\end{acknowledgments}

\end {document}